\RequirePackage{silence}
\documentclass[reprint,amsmath,amssymb,aps,pra,superscriptaddress,
    citeautoscript,
    floatfix]{revtex4-2}
\usepackage[utf8]{inputenc}
\usepackage[english]{babel}
\usepackage{hyperref}
\usepackage{bm}
\usepackage{hyperref}
\usepackage{physics}
\usepackage{tabularx}
\usepackage[page]{appendix}
\usepackage{nicefrac}
\usepackage[per-mode=power, 
            exponent-product = \times, 
            inter-unit-product = \ensuremath{{\cdot}}, 
            tight-spacing = true,
            number-unit-product=~,
            parse-numbers=false,]{siunitx}
\AtBeginDocument{\RenewCommandCopy\qty\SI}
\ExplSyntaxOn
\msg_redirect_name:nnn { siunitx } { physics-pkg } { none }
\ExplSyntaxOff
\usepackage{multibib}
\usepackage[dvipsnames]{xcolor}
\usepackage{booktabs}
\usepackage{float}
\usepackage{graphicx}
\usepackage[capitalise]{cleveref}
\usepackage{mhchem}
\usepackage{tabularx}
\usepackage{ulem}
\usepackage{enumitem}
\usepackage[
    protrusion=true,
    activate={true,nocompatibility},
    final,
    tracking=true,
    kerning=true,
    spacing=true,
    factor=1000]{microtype}
\hypersetup{
 colorlinks=true, 
 linkcolor=blue, 
 citecolor=blue,        
 filecolor=blue,      
 urlcolor=blue}
\usepackage[raggedright]{titlesec}
\usepackage{caption}
\titleformat{\section}{\raggedright\bfseries\large}{\thesection}{0pt}{}[]
\titleformat{\subsection}{\raggedright\bfseries}{\thesection}{0pt}{}[]
\titlespacing*{\section}{0em}{0.75em}{0pt}
\titlespacing*{\subsection}{0em}{0.75em}{0pt}
\renewcommand{\textcite}[1]{\citenum{#1}}
\newcolumntype{Y}{>{\centering\arraybackslash}X}
\Crefformat{appendix}{Supplementary Information S.#2#1#3}

\begin{document}
\setcitestyle{comma,sort&compress}
                                      
\author{Pradyoth~Shandilya}
\thanks{These authors contributed equally}
\affiliation{University of Maryland, Baltimore County, Baltimore, MD, USA}
\author{Shao-Chien~Ou}
\thanks{These authors contributed equally}
\affiliation{Joint Quantum Institute, NIST/University of Maryland, College Park, USA}
\author{Jordan~Stone}%
\affiliation{Joint Quantum Institute, NIST/University of Maryland, College Park, USA}
\affiliation{Microsystems and Nanotechnology Division, National Institute of Standards and Technology, Gaithersburg, USA}
\author{Curtis~Menyuk}%
\affiliation{University of Maryland, Baltimore County, Baltimore, MD, USA}
\author{Miro~Erkintalo}%
\affiliation{Department of Physics, University of Auckland, Auckland, New Zealand}
\affiliation{The Dodd-Walls Centre for Photonic and Quantum Technologies, Dunedin, New Zealand}
\author{Kartik~Srinivasan}%
\email{kartik.srinivasan@nist.gov}
\affiliation{Joint Quantum Institute, NIST/University of Maryland, College Park, USA}
\affiliation{Microsystems and Nanotechnology Division, National Institute of Standards and Technology, Gaithersburg, USA}
\author{Gr\'egory~Moille}%
\email{gregory.moille@nist.gov}
\affiliation{Joint Quantum Institute, NIST/University of Maryland, College Park, USA}
\affiliation{Microsystems and Nanotechnology Division, National Institute of Standards and Technology, Gaithersburg, USA}
\date{\today}

\title{All-Optical Azimuthal Trapping of Dissipative Kerr Multi-Solitons for\\Relative Noise Suppression}
\begin{abstract}
    Temporal cavity solitons, or dissipative Kerr solitons (DKS) in integrated microresonators, are essential for deployable metrology technologies. Such applications favor the lowest noise state, typically the single-DKS state where one soliton is in the resonator. Other multi-DKS states can also be reached, offering better conversion efficiency and thermal stability, potentially simplifying DKS-based technologies. Yet they exhibit more noise due to relative soliton jitter, and are usually not compatible with targeted applications. We demonstrate that Kerr-induced synchronization, an all-optical trapping technique, can azimuthally pin the multi-DKS state to a common reference field. This method ensures repetition rate noise independent of the number of solitons, making a multi-DKS state indistinguishable from a single-DKS state in that regard, akin to trapped-soliton molecule behavior. Supported by theoretical analysis and experimental demonstration in an integrated microresonator, this approach provides metrological capacity regardless of the number of cavity solitons, benefiting numerous DKS-based metrology applications.

\end{abstract}

\maketitle
\begin{figure}[t]
    \begin{center}
    \includegraphics{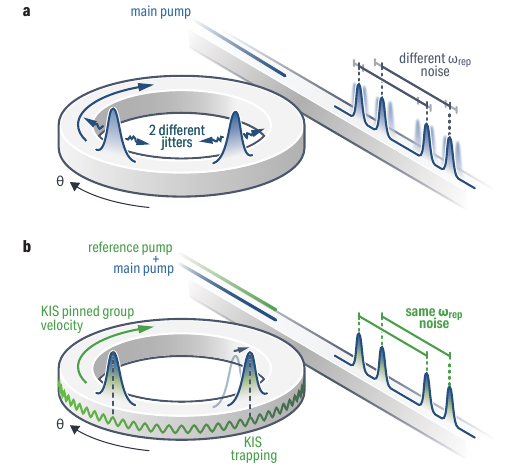}
    \caption{\label{fig:1}%
    Impact of Kerr-induced synchronization (KIS) on multi-soliton states. \textbf{a} Single-pump case where two dissipative Kerr solitons (DKSs) live in the cavity. They are both in phase since they are pumped by the same continuous wave laser. However, since they are not bound to one another, their jitters are independent, i.e., there is a relative jitter between the DKSs. This results in an output pair of pulse trains carrying independent repetition rate noise, yielding a noisier optical frequency comb than in the single-DKS case. %
    \textbf{b} Kerr-induced synchronization enables the phase locking of a DKS to the intracavity reference field, produced by sending another weak reference continuous wave pump laser into the microring. In the multi-DKS state, we show that both DKSs synchronize to the same common reference field, which pins their azimuthal positions, azimuthally trapping them and resulting in the suppression of the relative jitter. As a result, the repetition rate noise from the output pulse train now exhibits the same noise characteristics as the single-DKS state.
    }
    \end{center}
\end{figure}

The prospect of generating cavity solitons (CSs) was extensively theoretically explored in the 1990s~\cite{RosanovJOSAB1990, DAlessandroPhys.Rev.Lett.1991, TlidiPhys.Rev.Lett.1994, SpinelliPhys.Rev.A1998}, with the focus being on spatially diffractive resonators where the solitons would manifest themselves as self-localized, individually addressable spots on a two-dimensional spatial plane~\cite{LugiatoPhys.Rev.Lett.1987}. Research on spatial CSs targeted the realization of all-optical memories~\cite{CoulletChaos2004} and optical computing~\cite{MaggipintoPhys.Rev.E2000}, leveraging the ability to controllably trap the solitons in specific spatial locations, with experimental demonstrations achieved in semiconductor microcavities~\cite{BarlandNature2002, PedaciAppliedPhysicsLetters2006, GutlichChaos2007}. Despite such results, the limit in cavity size to house a large number of solitons, together with different spurious effects (cavity defects, free carrier processes, etc.) hindered robust development towards market applications. \\
\indent In part to overcome issues relating to \textit{spatial} CSs, the early 2010s saw significant efforts to explore \textit{temporal} CSs: pulses of light that can persist in passive waveguide resonators. With the object of addressing similar applications as their spatial counterparts, temporal CSs were first experimentally demonstrated in a macroscopic fiber ring resonator~\cite{LeoNaturePhoton2010}, and it was shown that they could be trapped in the time domain by appropriately modulating the coherent field driving the resonator~\cite{JangNatCommun2015, ErkintaloJ.R.Soc.N.Z.2022, EnglebertNat.Phys.2023}. While such trapping has enabled a temporal CS-based optical buffer operating at 10 GHz~\cite{JangOpt.Lett.OL2016}, technical requirements to sustaining the solitons make the application impractical as of today.\\
\indent Temporal CSs have also been realized in integrated optical microresonators, where they are commonly referred to as dissipative Kerr solitons (DKSs)~\cite{kippenberg_dissipative_2018}. In that context, the solitons have enabled the creation of optical frequency combs (OFCs), unlocking altogether new applications such as integrated optical clockworks~\cite{NewmanOptica2019,MoilleNature2023}, and low-noise microwave generation~\cite{KudelinNature2024,SunNature2024}, while maintaing compatibility with low size, weight, power and cost (SWAP-C) requirements~\cite{SternNature2018b, LiuNatCommun2021}.
Therefore, temporal DKS trapping, beyond applications originally envisaged for spatial DKSs, are of use in OFC systems. For example, it can be used to discipline the repetition rate for improved noise performance~\cite{WengPhys.Rev.Lett.2019a}. Yet, established methods relying on direct modulation of the monochromatic field driving the resonator or the cavity detuning~\cite{ErkintaloJ.R.Soc.N.Z.2022, EnglebertNat.Phys.2023,WengPhys.Rev.Lett.2019a} are not well suited for on-chip microring resonators since the DKS repetition rate is too high to be compatible with electronics that create the trapping potential. Hence, an all-optical technique is preferable, which can be achieved using bichromatic pumping~\cite{TaheriEur.Phys.J.D2017,ToddPhys.Rev.A2023}.  
To minimise the repetition rate noise, it is further preferable to operate with large bichromatic modal spacing, hence maximizing the resultant optical frequency division (OFD) factor. \\
\indent Recently, a novel scheme known as Kerr-induced synchronization (KIS) has been experimentally demonstrated~\cite{MoilleNature2023, WildiAPLPhotonics2023}, whereby a reference laser whose frequency is tuned close to a comb tooth makes it possible to synchronize the DKS comb to the reference, corresponding to phase trapping in the azimuthal domain~\cite{MoilleNature2023}. KIS efficiency can be further improved through on-resonance comb-tooth operation~\cite{MoilleNature2023,MoillearXiv2024} at a cavity mode close to the phase-matched dispersive wave frequency~\cite{AkhmedievPhysRevA1995,CherenkovPhys.Rev.A2017, BraschScience2016}. While the efficacy of this scheme has been demonstrated for single-DKS operation, its metrological impact on multi-DKS states remains completely unstudied. Addressing this shortcoming is particularly important, given that multi-DKS states have been shown to exhibit  higher conversion efficiency~\cite{JangOpt.Lett.OL2021} and enhanced robustness against thermal effects~\cite{GuoNat.Phys.2017} than single-DKS states. Yet, multi-DKS states are usually not desirable since they exhibit a relative jitter between each DKS [\cref{fig:1}a], leading to higher microcomb repetition rate noise~\cite{BunelAPLPhotonics2024}. Such multi-soliton relative jitter may be mitigated in the soliton-molecule regime, where two or more solitons interact directly with one another~\cite{GordonOpt.Lett.OL1983}, potentially in a long-range fashion through their tails~\cite{SmithOpt.Lett.OL1989}. Soliton molecules rely on the asymptotically stable balance between attractive and repulsive interactions and have been demonstrated in bulky fiber systems~\cite{MitschkeOpt.Lett.OL1987, StratmannPhys.Rev.Lett.2005} and crystalline millimeter cavities~\cite{YiNatCommun2018a, WengNatCommun2020a}. However, various sources of noise can still influence the solitons' characteristics (and hence mutual interactions), leaving residual relative jitter.~\cite{KrolikowskiOpt.Lett.OL1998}. In small resonators, molecule formation has relied on modal interaction (defect or avoided mode crossing) for different solitons to interact, resulting in their binding into so-called soliton crystals~\cite{ColeNaturePhoton2017}. In their more robust form, this leads to perfect soliton crystal formation~\cite{HeLaserPhotonicsRev.2020a} akin to cnoidal waves~\cite{QiOpticaOPTICA2019}. However, this sturdiness comes at the cost of a frequency comb whose frequency spacing is large (multiple times that of a single soliton state), making it impractical for repetition rate detection. Additionally, the soliton crystal behavior does not prevent the entire molecule from being sensitive to noise, particularly thermo-refractive noise, which is a dominant noise source in small microring resonators~\cite{StonePhys.Rev.Lett.2020, DrakeNat.Photonics2020, Moille2024_arXivTRN}. 
 To this extent, it is paramount to determine if a coherent temporal DKS trapping compatible with integrated photonics technology can exist, enabling predictable and deterministic low-noise microcomb operation independent of the microring design and multi-DKS state.\\
\indent In this work, we demonstrate that KIS enables relative azimuthal pinning of the different DKSs in an integrated silicon nitride (\ce{Si3N4}) microresonator through the shared reference intracavity field. Experimentally, we demonstrate that the same repetition rate noise can be measured---which is consistent with the optical division of the two pumps' noise---regardless of the number of DKSs present in the cavity. Our work highlights the metrological capacity of transforming a multi-DKS into low-noise coherent state through Kerr-induced synchronization.

\begin{figure}[t]
    \centering
    \includegraphics{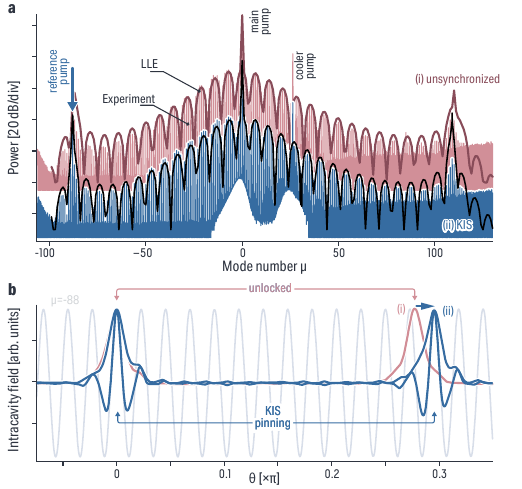}
    \caption{\label{fig:2}%
    The multi-DKS comb envelope with and without KIS. \textbf{a} Experimentally measured octave-spanning optical spectra of a 2-DKS state obtained outside of synchronization (i. red) and once in the KIS regime (ii. blue). The envelope in each case represents the LLE simulation of the system for its respective synchronization regime. %
    \textbf{b} Azimuthal domain study of the 2-DKS system obtained from the LLE simulations, with the trailing DKS referenced to $\theta=0$. In grey, the background intensity modulation from the reference field at $\mu_s=-88$ is shown, which in the synchronization regime azimuthally pins the leading DKS onto the same grid as the trailing one (blue), resulting in a position shift in comparison to the unsynchronized soliton case (red). This shift explains the modification of the interference pattern observed in the comb spectra. %
    }
\end{figure}


\section*{Results}

\subsection*{Experimental demonstration and empirical model of KIS-mediated locking of a multi-DKS state}
First, we aim to experimentally investigate the trapping of a multi-DKS state. To do so, we use a similar \ce{SiO2}-embedded \ce{Si3N4} microring resonator as in ref.~\textcite{MoilleNature2023}, and described in more detail in the Methods section. We use a main pump at \qty{\omega_\mathrm{0}/2\pi \approx 282.5}{\THz} (\qty{1061.9}{nm}) which, given the resonator dispersion, generates an octave-spanning frequency comb with dual dispersive waves at modes $\mu=-88$ and $\mu=109$ relative to the main pump. %
With this system, different multi-DKS states can be obtained [\cref{fig:2}a]. For the current experimental demonstration, we select a two-DKS state, while other states that can be reached will be discussed later. To synchronize the DKS, we use a reference laser that is injected close to the DKS comb line (within a KIS locking window) around the dispersive wave at mode $\mu = -88$ \qty{194.6}{\THz}. We verify synchronization through the repetition rate ($\omega_\mathrm{rep}$) disciplining~\cite{MoilleNature2023} and set the reference laser frequency so that the same $\omega_\mathrm{rep}$ is exhibited both in and out of synchronization, for a fair comparison (this occurs at the center of the KIS locking window). Our experiments show that the two-DKS state differs when in- and out-of-synchronization [\cref{fig:2}a], in particular in the interference pattern between the two DKSs that leads to modulation of the comb spectral envelope. It is worth pointing out that this phenomenon is reversible from and to the KIS regime, since once the reference pump is turned off the comb envelope pattern returns to the unsynchronized state, highlighting the reconfigurable aspect of this all-optical trapping. \\

\indent The repetition rate of the OFC that we obtain (\qty{\omega_\mathrm{rep}/2\pi\approx1}{\THz}) is too high for direct temporal detection of the azimuthal trapping of the DKSs under KIS. To gain insight on the nature of such synchronization, we use the modifed Lugiato-Lefever equation (mLLE), describing the DKSs' dynamics. The LLE was first developed for spatial DKSs~\cite{LugiatoPhys.Rev.Lett.1987}, then adapted to the temporal domain~\cite{HaeltermanOpticsCommunications1992}, and then formulated for the specific case of DKS comb generation in microresonators~\cite{ChemboPhys.Rev.A2013, CoenOpt.Lett.OL2013}, highlighting the similar physics, including trapping. Our system can be described by a mLLE that includes multiple driving forces~\cite{TaheriEur.Phys.J.D2017}, written as:
\begin{equation}
    \begin{split}
    \label{eq:MLLE}
    \frac{\partial \psi(\theta,t)}{\partial t} =& -(1 + i\alpha_p)\psi+ i|\psi|^2\psi\\ 
    +&i\sum_{\mu}\mathcal{D}(\mu)\tilde{\Psi}(\mu,t) \mathrm{e}^{i\mu \theta} 
    + F_\mathrm{p} \\
    +& F_\mathrm{ref} \exp\Bigl[ i\Bigl(\alpha_\mathrm{ref} - \alpha_p + \mathcal{D}(\mu_s)\Bigr)t + i\mu_s\theta\Bigr]
    \end{split}
\end{equation}
with a normalization similar to ref.~\textcite{ChemboPhys.Rev.A2013} that is relative to the total losses in the cavity $\kappa$. Here, $\psi(\theta,t)$ is the intracavity field and $\tilde{\Psi}(\mu,t) = \int_{\pi}^{\pi} \psi(\theta,t)\mathrm{e}^{-i\mu \theta}$ is its Fourier transform, $\theta$ and $\mu$ are the azimuthal coordinate and the azimuthal mode number, respectively, $t$ is the normalized time, $\alpha_p$ and $\alpha_{\text{ref}}$ are the normalized detuning of the primary and reference pumps respectively, $F_\mathrm{p}$ and $F_\mathrm{ref}$ are the external drive amplitudes of the primary and reference pumps, and $\mathcal{D}$ is the normalized integrated dispersion of the cavity in the $\mu$ space. More details regarding the simulation parameters are described in the Methods. 

We numerically solve the mLLE and verify that the resulting comb spectra matches against the experiment [\cref{fig:2}a]. We can then extract the azimuthal profile of the two-DKS state both in- and out-of-synchronization [\cref{fig:2}b]. In the unsynchronized case, the background modulation from the dispersive wave is small, and insufficient to trap the two DKSs into a molecule~\cite{WangOptica2017, SkryabinOpt.Express2017, WangOptica2017,ToddPhys.Rev.A2023}. In contrast, in the KIS regime, the background modulation is large enough to lock the two DKSs at grid points that are defined by the modulation. This modulation arises from the beating between the main and reference lasers that are separated by $\mu = -88$ mode numbers. This behavior is reflected in experiment by the change in the interference pattern observed in the comb envelope~[\cref{fig:2}a].

\subsection*{Linear stability analysis}
\begin{figure}[t]
    \centering
    \includegraphics{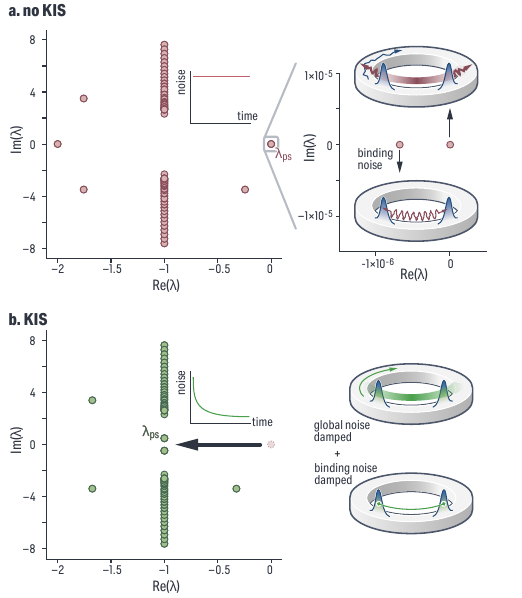}
    \caption{\label{fig:3}
    Dynamical spectrum of the linearized Lugiato-Lefever equation describing the multi-DKS system, for a two-soliton state in the cavity. %
    \textbf{a} Out of synchronization, the position-shifting eigenvalues (also known as the neutral or Goldstone modes) for the global two-DKS system and their relative position are close to $\lambda_{ps}=0$, highlighting both a sensitivity to intra-cavity noise and the independent jitter of the two DKSs in the multi-DKS state. 
    \textbf{b} In the KIS regime, the two $\lambda_{ps}$ of interest migrate to $-1$. As ref.~\cite{Moille2024_arXivTRN} highlights, when susceptibility to noise is determined by the eigenvalue(s) with the largest real parts, which in this case are of the order of the photon lifetime. Hence, the two-DKS KIS-mediated molecule has a stability in the presence of noise that is comparable to the stability of a single DKS. We note that the other eigenvalue whose real part is at $-2$ in \textbf{a} also migrates to $-1$, hybridizing with the $\lambda_\mathrm{ps}$ and causing the split in the imaginary part we see in KIS.
    }
\end{figure}

    The above empirical demonstration highlights the coherent trapping aspect of the multi-DKS state to the same reference field, yet does not fully elucidate the solitons' relative dynamical stability. We can further our understanding of the system by performing a linear stability analysis of the DKS solutions in~\cref{eq:MLLE}~\cite{MaggipintoPhys.Rev.E2000,FirthPhys.Rev.Lett.1996, ErkintaloJ.R.Soc.N.Z.2022, ToddPhys.Rev.A2023,Moille2024_arXivTRN,GasmiPhys.Rev.A2023a}. Contrary to most linear analysis studies that focus solely on the single-soliton solution, we focus here on the two-DKS state, this analysis enables us to retrieve the stability of their common group velocity along with that of their relative jitter, providing new insights into the noise properties of multi-soliton states.  After a convenient change of variable further detailed in ref.~\cite{Moille2024_arXivTRN}, we linearize the equation following ref.~\textcite{WangJ.Opt.Soc.Am.B2018a} such that: 
\begin{equation}\label{eq:linearized}
    \frac{\partial \Delta\psi{(\theta, t)}}{\partial t} = \mathcal{L}\left[\psi_0{(\theta, t)}\right]\Delta\psi{(\theta, t)},
\end{equation}
\noindent where $\psi_0(\theta, t)$ is the temporal stationary DKSs solution obtained by solving~\cref{eq:MLLE} with its left-hand side set to zero and setting the correct main pump power and detuning to get the targeted intracavity state~\cite{QiOptica2019}. 
$\mathcal{L}$ is a linearized operator, and $\Delta\psi(\theta,t)$ is a perturbation of the stationary solution $\psi_0(\theta,t)$. A perturbation can generally be decomposed through a linear superposition of eigenfunctions $v_n(\theta)$ of $\mathcal{L}$ with associated eigenvalues $\lambda_n$, so that after a time $\Delta t$, one can express the perturbation to the intracavity field as: 

\begin{align}\label{eq:perturbation}
    \Delta \psi(\theta, t_0 + \Delta t) &= \sum_n \exp[\lambda_n \Delta t]a_n v_n(\theta).
\end{align}
Hence, from the study of $\lambda_n$, we can conclude if the perturbation is amplified ($\Re(\lambda_n)>0$), persists ($\Re(\lambda_n)=0$), or is damped ($\Re(\lambda_n)<0$). Although the linearized operator presents as many eigenvalues as the number of modes used to model the system~[\cref{fig:3}], we are interested in one in particular -- the so-called position-shifting eigenvalue ($\lambda_{ps}$), also called the neutral mode or Goldstone mode~\cite{FirthPhys.Rev.Lett.1996, MaggipintoPhys.Rev.E2000}---which is responsible for group velocity noise. Since we study a multi-DKS state, here for the sake of simplicity a two-DKS state, the eigenvector projection results in studying the projected basis of the two-DKS system. This yields two values for $\lambda_{ps}$ of interest: one for the global multi-DKS behavior and one for the two DKSs' relative motion~\cite{VladimirovPhys.Rev.A2021}, with the former informing us about the repetition rate noise and the latter about the multi-DKS relative jitter.  
In the single pump case [\cref{fig:3}a], both $\lambda_{ps}$ values exist around zero, highlighting that any intracavity noise will persist and that the two-DKSs are weakly bound, so that their group velocity noise is independent and uncorrelated. In striking contrast, in KIS [\cref{fig:3}b], the two $\lambda_{ps}$ values have migrated toward $-1$, corresponding to the maximum damping of the intracavity noise at a photon-lifetime rate $\tau_\mathrm{phot} = 1/\kappa$ that is consistent with the initial normalization in~\cref{eq:MLLE}. Such maximization of the damping occurs when the reference laser is at the center of the KIS spectral window, namely at the comb tooth frequency of the unsynchronized DKS. For different reference frequencies within the KIS window, $\lambda_{ps}$ will exhibit continuous values from $-1$ (center of KIS) to $0$ (edge of KIS)~\cite{Moille2024_arXivTRN}. We note that a hybridization between $\lambda_\mathrm{ps}$ and the other eigenvalues whose real part are originally at $-2$ and migrate to $-1$ leads to the splitting of the imaginary part of the eigenvalue in KIS, which does not impact the noise damping properties. Since both the global and relative position shifting eigenvalues are damped at the same rate, the relative position in time of the two-DKS system will exhibit correlated noise that is only limited by the photon lifetime, similar to the single-DKS case~\cite{Moille2024_arXivTRN}. Therefore, using a single- or multi-DKS while synchronized to a reference pump results in the same outcome from the linear stability analysis, and hence the noise properties of the measured repetition rate are expected to be the same. This suggests that under KIS, the multi-DKS state becomes compatible with low noise metrology applications.
                        
\subsection*{Noise performance comparison between single- and multi-DKS in KIS}
\begin{figure}[t]
    \centering
    \includegraphics{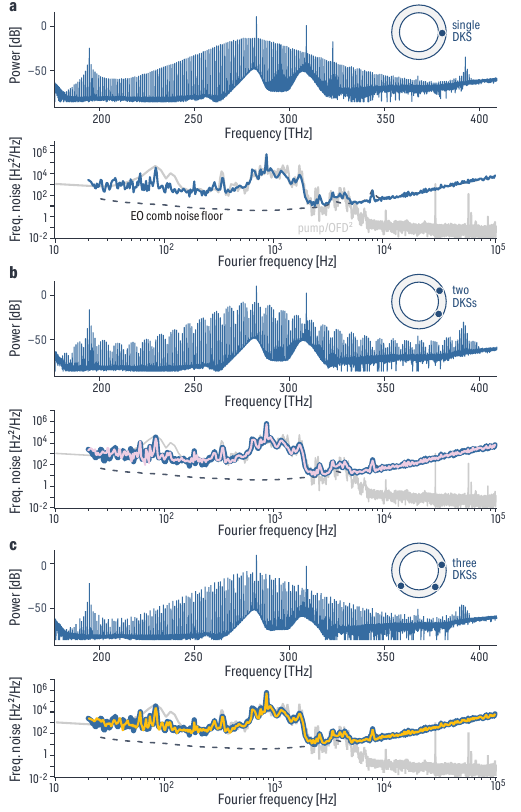}
    \caption{\label{fig:4}%
    Noise characterization of the repetition rate ($\omega_\mathrm{rep}$) in KIS of \textbf{a} a single-DKS state, \textbf{b} two-DKS state and \textbf{c} three-DKS state. While the spectra look different due to the different number of DKSs living inside the microring cavity (top inset for each plot), the frequency noise spectral density remains the same as in the single-DKS state (blue trace in the bottom of a, b and c), either for a two-DKS state (b, pink) or three-DKS state (c, yellow). As expected from KIS operation, the repetition rate noise corresponds to the two uncorrelated pumps' noise optically frequency divided by a factor $OFD=\mu_s^2$ (grey trace in a, b and c). The noise floor of the electro-optic comb (dashed curve), which enables measurement of \qty{\omega_\mathrm{rep}\approx1}{\THz}, prevents the repetition rate noise from overlaying on the frequency-divided pump noise beyond \qty{5}{\kHz}. Unsynchronized repetition rate noise for each DKS state is not measured since it is higher than the detection limit of our phase noise analyzer, mostly due to the high thermorefractive noise of the microcomb, which does not play a role when the system is synchronized, as discussed in Ref.~\cite{Moille2024_arXivTRN}.
    }
\end{figure}

In this final section, we proceed to measure the power spectral density of the repetition rate frequency noise $S_\mathrm{rep}(f)$ for synchronized single- and multi-DKS states. Using the same exact device as the one presented in~\cref{fig:2}, we proceed to generate different DKS states. In order to do so, we slightly tune both the main pump power (within \qty{10}{\mW}) and main pump detuning to adiabatically land on a multi-DKS state, from which we determine the number and spacing of solitons by fitting to the observed spectral envelope. The repetition rate is measured using an electro-optic comb (EOcomb) apparatus described in more detail in the Methods section.%

In the case of the single-DKS [\cref{fig:4}a], where the comb spectrum has the signature smooth envelope, the obtained $S_\mathrm{rep}(f)$ exhibits the same noise as the main and reference pump noise frequency divided by the optical frequency division factor $OFD=\mu_s^2 = 88^2 = 7744$, as expected from previous study~\cite{MoilleNature2023,WildiAPLPhotonics2023}, since the characteristic frequency noise of the pumps is well-reproduced for frequencies below \qty{5}{\kHz} after accounting for the OFD. The EOcomb apparatus which lets us frequency translate the large DKS $\omega_\mathrm{rep}$ to a detectable bandwidth limits our noise floor to about \qty{10^2}{\Hz\squared\per\Hz} at a Fourier frequency of \qty{2}{\kHz} and \qty{10^4}{\Hz\squared\per\Hz} at a Fourier frequency of \qty{100}{\kHz}. After generating different multi-DKS states, either a two-DKS microcomb [\cref{fig:4}b] or a three-DKS one [\cref{fig:4}c], the obtained $S_\mathrm{rep}(f)$ presents the same characteristics as the single-DKS state, with the same optically frequency divided noise from the pumps, despite the multiple cavity solitons traveling within the resonator. Such noise measurements validate the metrological capacity of multi-DKS states in the KIS regime, demonstrating that the azimuthal trapping of the DKSs by the reference field is strong enough to provide a single repetition rate noise, consistent with the theoretical linear stability described earlier.

\section*{Discussion}
To conclude, we have demonstrated that Kerr-induced synchronization enables trapping of general multi-DKS states, which leads to an on-demand low-noise coherent state, where all the pulses are trapped by the background modulation at a period $\mu_s$ defined by the mode at which the reference synchronizes the DKSs. Experimentally, the spectrum of a two-DKS state provides sufficient discrepancy in and out of synchronization to highlight the azimuthal trapping of the different DKSs to the $\mu_s$ reference modulation. This enables a direct comparison with pump phase-modulation trapping and provides an experimental demonstration of previous theoretical work where the background modulation period is similar to the DKS pulse width~\cite{ToddPhys.Rev.A2023}. We analyze the multi-DKS state and its Kerr-induced synchronization through a theoretical linear stability analysis of the mLLE, observing similar behavior as in the single-DKS state, which per definition is a signature of the molecular behavior. In the linearized system, the eigenvalue of interest $\lambda_{ps}$ has a real part that moves from zero (no intra-cavity noise damping out of synchronization) to $-1$ (damping at the photon decay rate at the center of the KIS window). In the multi-DKS case, the real part of both the global multi-DKS eigenvalue and the eigenvalue corresponding to the relative jitter between the DKSs is present at $-1$, demonstrating the relative noise suppression between each DKS. Finally, we show experimentally that the repetition rate frequency noise ends up being identical in a single-DKS or in a multi-DKS state, which is solely determined by the pump frequency noise and the optical frequency division factor $\mu_s^2$, as usual in KIS. Our work highlights that KIS brings metrological capacity to multi-DKS states through azimuthal trapping of the DKS and on-demand predictive low-noise operation. Since integrated frequency combs rely on low SWaP-C, the overall comb power remains largely limited, but can be mitigated by increasing the number of pulses in the cavity. Usually, this leads to a trade-off between either the increase of the repetition rate noise with the multi-DKS state order or an increase in the comb tooth spacing to harmonics of $\omega_\mathrm{rep}$ in the particular case of a multi-DKS state often referred to as a soliton crystal~\cite{ColeNaturePhoton2017}. Although control of the defects that can produce such soliton crystal states has been demonstrated with laser injection~\cite{TaheriNatCommun2022b}, our work instead shows that KIS enables a coherent azimuthal trapping of all the DKSs present in the cavity. We therefore  demonstrate that one can leverage the benefits of multi-DKS states without being metrologically limited or being forced to work with harmonics of the repetition rate. %
In addition, since arbitrary multi-state DKS orders exhibit the same noise performance as the single-DKS state, our work could be extended to larger resonator circumference (\textit{i.e.} lower $\omega_\mathrm{rep}$ combs) where much more pulses could be fit inside the resonator, significantly increasing the otherwise poor conversion efficiency of the microcomb operation without sacrificing noise performance.

                          
%

                                        
\section*{Methods}
\subsection*{Microring resonator design}
The photonic chips were fabricated following the process presented in ref~\textcite{MoilleNature2023} in a commercially available foundry. We use a silicon nitride (\ce{Si3N4}) microring resonator embedded in \ce{SiO2} with an outer ring radius \qty{R=23}{\um} and ring width \qty{RW=850}{\nm}. A bus waveguide (width \qty{W_\mathrm{wg}=460}{\nm}) wrapped around the ring in a pulley-like fashion~\cite{MoilleOpt.Lett.2019} with coupling length \qty{L_\mathrm{c}=17}{\ um} and gap \qty{G=600}{\nm} enables critical coupling at both the main pump frequency \qty{\omega_\mathrm{0}/2\pi \approx 282.5}{\THz} (\qty{1061.9}{nm}) main pump and the reference pump frequency \qty{\omega_\mathrm{ref}/2\pi\approx194.6}{\THz} (\qty{1549.6}{nm}). The reference is selected to be at the low frequency dispersive wave (DW) at $\mu_s = -88$ comb teeth away from the main pump. To access the soliton state, we use about \qty{150}{\mW} of on-chip main pump power while temperature stabilizing the resonator through a cooler-pump~\cite{ZhangOptica2019, ZhouLightSciAppl2019} with about \qty{250}{\mW} of on-chip power in a counter-propagating and cross-polarized (transverse magnetic) mode with respect to the main pump (to minimize nonlinear interaction). 

\subsection*{Simulation parameters}
The simulation result of the Lugiato-Lefever equation (LLE) presented in~\cref{fig:2} used the same geometric parameter of the microring resonator as {presented above, with the dispersion parameter $\mathcal{D}$ in the LLE obtained via accurate finite element method (FEM) modelling}~\cite{MoilleNature2023}. The simulated intrinsic and quality factors are set to $Q_i = 1\times10^6$ and $Q_c=1\times10^6$ respectively. The detuning of the pump is set to be \qty{\bigl(\omega(\mu = 0) - \omega_0 \bigr)/2\pi=1.25\times10^9}{\GHz}, with the main pump frequency \qty{\omega_0 / 2\pi \approx 282.5}{\THz}, while the round trip time of the DKS is \qty{t_r = 2\pi/\omega_\mathrm{rep} \approx 1}{\ps}, yielding a normalized parameter $\alpha = 4.4$. The main pump is set at an in-waveguide power of \qty{P_\mathrm{in}=160}{\mW}, and the effective nonlinearity \qty{\gamma=3.2}{\per \W\per\m} from FEM calculations, resulting in a normalized driving force of $F_0 = 2.78$ and a normalized time $\tau = 1.7\times10^9 t$

\subsection*{Repetition rate detection}
In order to measure the close to \qty{1}{\THz} repetition rate of the microcomb, we spectrally translate two adjacent comb teeth close to one another to measure their beats, similar to refs.~\textcite{StonePhys.Rev.Lett.2020, MoilleNature2023}. To perform such spectral translation, we use an electro-optics (EO) comb consisting of two phase modulators driven at \qty{\omega_\mathrm{EO} = (17.839 \times 2\pi)}{\GHz}, enabling two adjacent microcomb teeth at \qty{\approx 271.5}{\THz} and \qty{\approx 270.5}{\THz} to be spectrally translated by $N_\mathrm{EO}=56$ EOcomb lines. Using a band-pass grating optical spectral filter with a \qty{0.1}{\nm} Gaussian shape full-width half maximum pass-band, one can measure the beat note at \qty{\omega_\mathrm{beat} =  22.75}{\MHz} (\qty{\pm10}{\kHz}), letting us determine the DKS repetition rate \qty{\omega_\mathrm{rep} = N_\mathrm{EO}\omega_\mathrm{EO} + \omega_\mathrm{beat} = 999.00675}{\GHz}~\qty{\pm10}{\kHz}. $\omega_\mathrm{beat}$ is detected by a \qty{50}{\MHz} bandwidth avalanche photodiode and processed with a phase-noise analyzer to study the repetition rate frequency noise.%

\section*{Acknowledgments}
S-C.O, K.S., and G.M acknowledge partial funding support from the Space Vehicles Directorate of the Air Force Research Laboratory and the NIST-on-a-chip program of the National Institute of Standards and Technology. P.S and C.M. acknowledges support from the Air Force Office of Scientific Research (Grant No. FA9550-20-1-0357) and the National Science Foundation (Grant No. ECCS-18-07272). M.E acknowledges financial support from the Marsden Fund of the Royal Society of New Zealand (Grant No. 23-UOA-071). G.M dedicates this work to T.B.M -- always.

\section*{Author contributions}
S-C.O. and G.M performed the experimental work. P.S. developed the theory and performed the simulations. J.S. helped with the metrology experiment and understanding. C.M., M.E, and K.S. contributed in the understanding of the physical phenomenon. G.M. supervised the project. G.M wrote the manuscript with the help of J.S, M.E and K.S. All the authors contributed and discussed the content of this manuscript.

\section*{Competing interests.}%
G.M., C.M. and K.S have submitted a provisional patent application based on aspects of the work presented in this paper.
\section*{Additional Information}
Correspondence and requests for materials should be addressed to G.M.
\section*{Data availability.}
The data that supports the plots within this paper and other findings of this study are available from the corresponding authors upon reasonable request.

\end{document}